# Electron Spin Resonance Shift and Linewidth Broadening of Nitrogen-Vacancy Centers in Diamond as a Function of Electron Irradiation Dose

Edwin Kim,[1] Victor M. Acosta,[2,3] Erik Bauch,[3,4] Dmitry Budker,[3,5] and Philip R. Hemmer[6,*]

[1] Ramtron International Corporation, 1850 Ramtron Drive, Colorado Springs, CO 80921, USA
[2] Hewlett-Packard Laboratories, 1501 Page Mill Road, Palo Alto, CA 94304, USA
[3] Department of Physics, University of California, Berkeley, CA 94720-7300, USA
[4] Technische Universität Berlin, Hardenbergstrasse 28, 10623 Berlin, Germany
[5] Nuclear Science Division, Lawrence Berkeley National Laboratory, Berkeley, CA 94720, USA
[6] Department of Electrical and Computer Engineering, Texas A&M University, College Station, TX 77843-3128, USA

**ABSTRACT** A high-nitrogen-concentration diamond sample was subject to 200-keV electron irradiation using a transmission electron microscope. The optical and spin-resonance properties of the nitrogen-vacancy (NV) color centers were investigated as a function of the irradiation dose up to $6.4 \times 10^{21}$ e$^-$/cm$^2$. The microwave transition frequency of the NV$^-$ center was found to shift by up to 0.6% (17.1 MHz) and the linewidth broadened with increasing electron-irradiation dose. Unexpectedly, the measured magnetic sensitivity is best at the lowest irradiation dose, even though the NV concentration increases monotonically with increasing dose. This is in large part due to a sharp reduction in optically-detected spin contrast at higher doses.

**KEYWORDS** Electron Spin Resonance, Nitrogen-Vacancy, Diamond Defect, Magnetometer, Sensitivity, Electron Irradiation, TEM

The nitrogen-vacancy (NV) center in diamond has been explored recently for many applications including quantum information,[1] magnetic sensors,[2,3,4,5] and subwavelength imaging.[6] Much of the NV utility is due to its optically detectable ground-state electron spin resonance. To achieve the best performance of magnetic-sensing devices utilizing ensembles of NV⁻ centers, a high concentration of the centers is desired.[7] This is achieved by either implanting nitrogen into pure diamond or by creating vacancies in nitrogen-rich diamond, followed by annealing to produce NV centers. Substitutional nitrogen atoms ($N_S$) that do not form NV centers are a source of spin dephasing, so it is important to optimize the conversion of $N_S$ to NV. One way to achieve this is with high-dose electron irradiation followed by annealing.[8] In this paper the effects of irradiation damage on the magnetic-sensing properties of the NV centers are explored.

Vacancies can be created using a variety of irradiating species, including electrons, neutrons, protons, and ions. Koike *et al* reported on the displacement threshold energy, $T_d$, of type-IIa natural diamond using a transmission electron microscope (TEM) for three principal crystallographic directions, [100], [110], and [111].[9] It was found that $T_d$ was 37–48 eV and the minimum incident-electron energy needs to be 180, 210, and 220 keV, respectively, for the [100], [111], and [110] orientations in order to form displacement-related defects. Steeds and co-workers demonstrated the creation of self-interstitials and carbon-carbon pairs along [100], using a 300 keV TEM.[10,11] Campbell and Mainwood predicted the radiation damage of diamond caused by both electron and gamma irradiation.[12] Recent work has focused on proton and electron irradiation on diamond, studying the converted NV⁻ and NV⁰ concentrations for optical-magnetometer applications[8] as well as NV⁻ formation using low-energy electrons.[13]

In this work, we used TEM irradiation of [100]-oriented nitrogen-rich type-Ib single-crystal bulk diamond. After irradiation, the vacancies are made mobile by annealing at approximately 700°C. The vacancies bind with a neutral substitutional nitrogen center, $N_S^0$, to form an $NV^-$ center as follows[14,15]: $N_S^0 + V^0 \rightarrow NV^0$ and $NV^0 + N_S^0 \rightarrow NV^- + N_S^+$. The latter reaction assumes that a second nitrogen center serves as an electron donor to enhance the fraction of negatively charged NV centers. Here the [100] orientation is chosen because it has a lower electron-energy threshold at room temperature for displacement of the carbon atoms compared to the other orientations, and this allows us to use a 200 keV TEM (JEOL JEM-2010).

It is now widely accepted that there are six valence electrons associated with the $NV^-$ center,[16,17] three from the dangling carbon bonds, two from the nitrogen, and one from a donor. As a result, the electronic ground state of the $NV^-$ center is a paramagnetic triplet state (S = 1). Figure 1 shows the NV center in diamond and a typical diamond fluorescence spectrum. Irradiated type-Ib bulk diamond presents two signature peaks at 575 nm and at 637 nm of $NV^0$ and $NV^-$, respectively [Fig. 1(b)]. In type-Ib diamond, most nitrogen impurities take the form of single substitutional nitrogens, which serve as electron donors.[18,19] For this reason, electron irradiation on type-Ib diamond, followed by annealing at temperatures above ~650° C, normally produces negatively-charged $NV^-$ centers.[20]

The diamond used in this experiment was a type-Ib single-crystal plate synthesized using the high-pressure, high-temperature (HPHT) method at Element Six with an initial substitutional-nitrogen concentration of $[N_S^0]$ = 65(10) ppm, as measured by infrared spectroscopy.[8] The irradiation was performed at room temperature with JEM-2010. Several spots on the sample were irradiated. For each spot, the electron beam was focused to a diameter of a few microns, and the exposure times for different spots were chosen to achieve doses ranging from $1.3 \times 10^{18}$ to $6.4\times10^{21}$ e⁻/cm². After all the spots were irradiated, the sample was annealed at 700° C for 1.5 hours in vacuum and cleaned using a combination of nitric and sulfuric acids.

We performed optical tests using scanning confocal microscopy. A 532 nm Nd:YAG laser illuminated the diamond sample through a 3.4 mm working-distance objective with numerical aperture of 0.8 and magnification of 100×. Dual-axis galvanometric mirrors were used to scan in the x-y plane of the sample and a fine z-scan was achieved by a piezo mount of the objective to adjust the focus (the depth of focus was calculated to be 1.2 μm). The fluorescence from the NV centers was detected either by a single-photon counting module (SPCM, PerkinElmer) using a silicon avalanche photodiode (APD) or a spectrometer. The scanned images before and after the annealing are shown in Fig. 1(c), where the absolute photon-count rate is reproducible within a factor of two between measurements. Prior to the annealing, the high-dose spots gave visible fluorescence, while the low-dose spots were not detected. After the annealing, a large increase in the APD count rate was seen, as expected due to formation of NV centers.

The photoluminescence spectrum for 532 nm excitation wavelength from each of the irradiated spots was measured at various depths before and after annealing using a Princeton Instruments SP-2150i spectrograph with a CCD detector (PIXIS:100, Princeton Instruments). Fluorescence spectra of each irradiated spot after annealing are shown in Fig. 1(d). The concentration of NV centers can be estimated from the integrated intensity of zero-phonon lines (ZPLs).[8] As the dose increases, the $NV^0$ fraction rises and the zero-phonon line of $NV^-$ shifts toward increasing wavelengths by up to 1.23 nm. The normalized $NV^0$ concentration appears to saturate at around $1\times10^{21}$ e⁻/cm². However, the normalized $NV^-$ concentration still appears to rise at the highest dose [Fig. 2(a)].

Spectra were recorded at various depths below the surface [Fig. 2(b)] to measure the NV formation profile.[21] The resulting distribution shows $NV^0$ and $NV^-$ features maximized at 5–10 µm below the surface, where the resolution in this direction is estimated at ~2.4 µm. The trajectories of 200 keV electrons in diamond were modeled with CASINO,[22] a Monte Carlo simulator, Fig. 2(c). The electron-energy distribution in bulk diamond is also shown in Fig. 2(d) as calculated with CASINO for 200 keV electrons. The CASINO simulations indicate that a 200 keV electron penetrates deeper than 100 µm. However, 200 keV electrons only retain enough energy to create vacancies within about 50 µm of the surface. This estimate is based on the 180 kV threshold for our [100] crystal, which means that vacancies can only be produced in the region where the electron still has above 90% of its incident energy (i.e., >180 kV).

The microwave-transition frequency of the $NV^-$ can be calculated from a Hamiltonian typical for a system with $C_{3v}$ symmetry: $H = g\beta_e \mathbf{B} \cdot \mathbf{S} + \mathbf{S} \cdot \mathbf{D} \cdot \mathbf{S}$ where $g$ is the electron $g$-factor, $\beta_e = 9.27 \times 10^{-24}$ J T$^{-1}$ is the Bohr magneton, $\mathbf{B}$ is the external magnetic field, $\mathbf{D}$ is the zero-field splitting tensor, and $\mathbf{S}$ is the electron spin. When the magnetic field ($B_z$) is applied along the quantization axis of the electron spin, this Hamiltonian has eigenfunctions: $|+1\rangle$, $|0\rangle$ and $|-1\rangle$ of $S_z$. If $B_z = 0$, there are three states, one of energy $E_0 = -\frac{2}{3}D$ and the other two of energy $E_{\pm 1} = \frac{1}{3}D$ where $D = 2.87$ GHz at room temperature [Fig. 3(a)].[23, 24]

In order to measure the spin properties of $NV^-$ centers, optically-detected electron spin resonance (ESR) measurements for the NV fluorescence were performed. The microwave signal from a PTS 3200 signal generator was amplified by a high-power microwave amplifier (ZHL-16W-43+, Mini-Circuits). The signal generator was set to operate in triggered mode in order to synchronize the data acquisition (DAQ) card for fluorescence detection. Microwaves were transmitted via a copper wire that was placed close to the irradiated area, as shown in Fig. 3(b). The microwave frequency was swept with no external static magnetic field applied. In Fig. 3(c), the ESR measurements are depicted for the different doses. The resonance position exhibits a shift towards higher resonance frequencies. As the dose goes up, the ESR frequency increases from 2.871 GHz by up to 17.1 MHz (0.6%), as shown in Fig. 3(d).

We note that ESR shifts were recently studied as a function of the sample temperature.[24,25,26] The nonlinear temperature dependence of zero-field splitting parameter $D$, $dD/dT$, was found to be proportional to the thermal expansion of the lattice ($dR/dT$, where R is the distance between two basal carbon atoms). We expect that the ESR frequency shift observed in Fig. 3(d) can also be correlated to dose-dependent changes in R. In fact, we observe a non-linear dependence of $dD/dD_e$ (the variation of $D$ depending on electron dose, $D_e$).

For the purpose of a magnetic sensor, type-1b diamond is of interest since it has a high concentration of substitutional nitrogen ($N_S$), and efficient conversion from $N_S$ to $NV^-$ can result in a high concentration of $NV^-$ centers.[8] The minimum detectable magnetic field of a dc magnetometer can be expressed by $\delta B \approx \hbar/[g\beta_e T_2^* (SNR)]$, where $\hbar$ is Planck's constant, SNR is the signal-to-noise ratio defined as the ratio of the signal intensity to the root-mean-square value of noise ($A_{signal}/\sigma_{noise}$), and $T_2^*$ is the effective inhomogeneous dephasing time.[27] The inhomogeneously broadened spin linewidth is given by $\Delta \nu = 1/(\pi \cdot T_2^*)$.

In Figs. 4(a) and (b), the normalized NV concentration and linewidth are plotted. For increasing dose, the linewidth broadens from 14.6 MHz to 17.1 MHz until the line finally disappears. As a result, $T_2^*$ declines by 15% for doses from $1.3 \times 10^{18}$ e$^-$/cm$^2$ to $2.6 \times 10^{20}$ e$^-$/cm$^2$. The dc magnetic sensitivity can be calculated using $T_2^*$ and the measured SNR taken from Figs. 4(b) and 4(c), respectively. As seen in Fig. 4(d), the minimum detectable field is optimized at the lowest irradiation dose, where it is 1.6 µT in a total integration time $t = 1.5$ ms. For higher doses, the sensitivity degrades substantially, even though the fluorescence intensity continues to increase.

This unexpected result is due to the fact that the SNR does not scale with shot noise, in large part because the spin contrast is not constant with increasing dose. If these effects could be eliminated, the shot noise prediction would give a $SNR \propto \sqrt{Nt/T_2^*}$ where $N$ is the number of NV centers assuming the total fluorescence is linear in NV concentration. This leads to the usual prediction for the shot-noise-limited minimum detectable field, $\delta B \approx \hbar / \left( g \beta_e \sqrt{NtT_2^*} \right)$.[7,28] As the product of $T_2^*$ and $N$ increases with dose throughout the range studied here [Figs. 4(a) and 4(b)], under ideal spin contrast $\delta B$ should decrease monotonically with increasing dose, contrary to experiment.

At present we do not know the reason for the sharp reduction of spin contrast with increasing irradiation dose at high doses. Since the ESR linewidth broadens only gradually with dose [Fig. 4(b)], it is unlikely that this could be due to insufficient driving amplitude of the microwave field. Furthermore, we still observe substantial NV⁻ emission at these high doses [see Fig. 1(d)], so it is unlikely that background fluorescence from other defects obscures the signal. We therefore tentatively conclude that the increased irradiation damage results in either decreased optical polarization or degraded spin-dependence of fluorescence (or some combination of the two).

In summary, after irradiation and annealing at 700°C, the photo-emission spectra from a type-Ib diamond were measured as a function of irradiation dose. Aside from increasing $NV^0$ and $NV^-$ concentrations with dose, the NV⁻ zero-field ESR frequency nonlinearly shifts upwards by 17.1 MHz, possibly due to diamond-lattice shrinking. At the highest doses, the accumulated irradiation damage results in the loss of the ESR signal. From the ESR spectra we computed the minimum detectable field of an optical magnetometer based on NV-doped diamond as a function of the electron irradiation dose. Surprisingly, we found the best sensitivity at the weakest electron irradiation dose, even as the $NV^-$ concentration continues to rise. Note that an ac magnetometer would have a much higher sensitivity than shown in Fig. 4(d) due to the fact that $T_2$ is much longer than $T_2^*$ and that classical noise often decreases rapidly with modulation frequency. Nonetheless, this work shows that considerable discrepancies exist between predicted and measured sensitivities for ensemble magnetometers based on NV-doped diamond, at least when using electron irradiation to create NVs from substitutional nitrogen in type-1b diamond. Thus it outlines important questions that should be addressed in future NV ensemble magnetometer research.

**Acknowledgements**. The authors are grateful to M. P. Ledbetter for valuable discussions and to D. Twitchen for providing the diamond sample. This work has been supported by NSF (awards # PHY-0855552 and ECCS-1202258), NIH, and the DARPA QuASAR program.

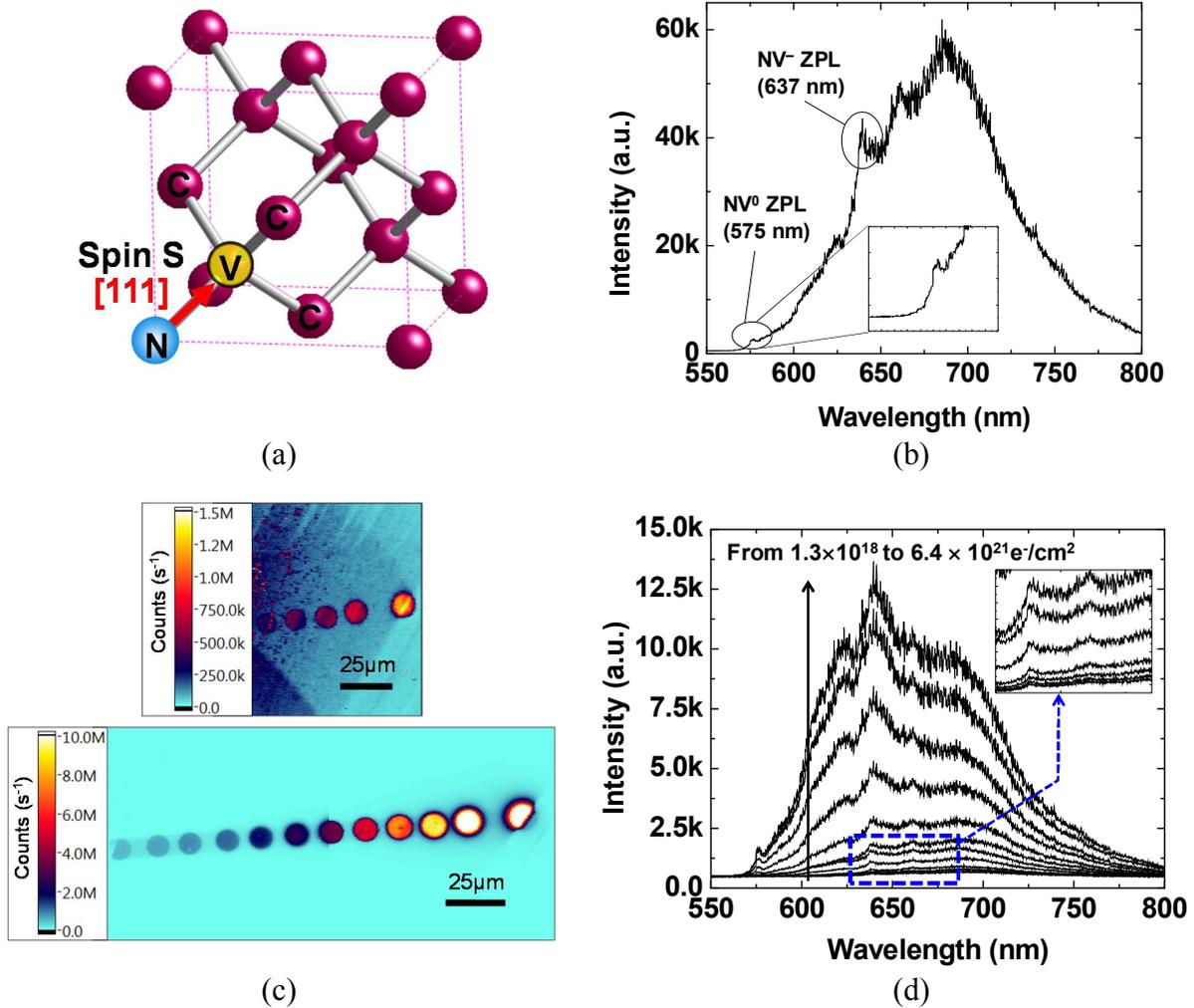

Figure 1. (a) The NV defect center is oriented along the [111] direction in the schematic diagram of the diamond lattice. (b) A typical room-temperature spectrum for the NV centers in a type-Ib diamond is displayed with the zero-phonon lines (ZPL) indicated. (c) Results of the confocal-microscopy scans of the irradiated diamond sample before annealing (top) and after annealing (bottom) with the optical power, $P_{op}$ = 1 mW. The spots correspond to the locations where the TEM beam was focused. The highest dose is on the right and the lowest dose is on the left on both figures. The fluorescence image before annealing is measured with no optical attenuator. The low-dose spots are not detected since their fluorescence is low compared to the background. In the scanned fluorescence images with ×0.005 light attenuator after annealing, the doses are 1.3, 2.6, 6.4×$10^{18}$; 1.3, 2.6, 6.4×$10^{19}$; 1.3, 2.6, 6.4×$10^{20}$; and 1.3, 2.6, 6.4×$10^{21}$, respectively, left to right. (d) The photoluminescence spectra for each dose obtained at the same optical power and integration time. As the dosage increases, it is observed that the $NV^0$ and $NV^-$ peaks also rise. The spectra for the low-dose spots are shown more clearly in the inset.

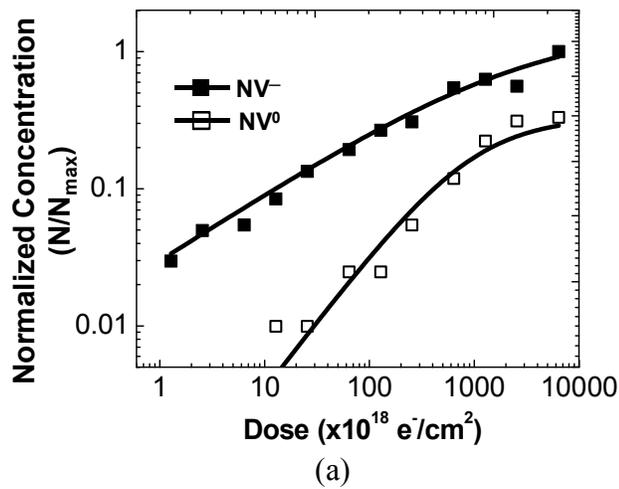
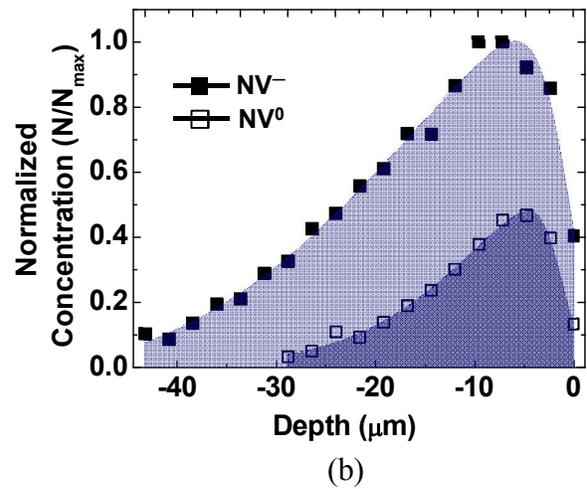

(a)　　　　　　　　　　　　　　　　　　(b)

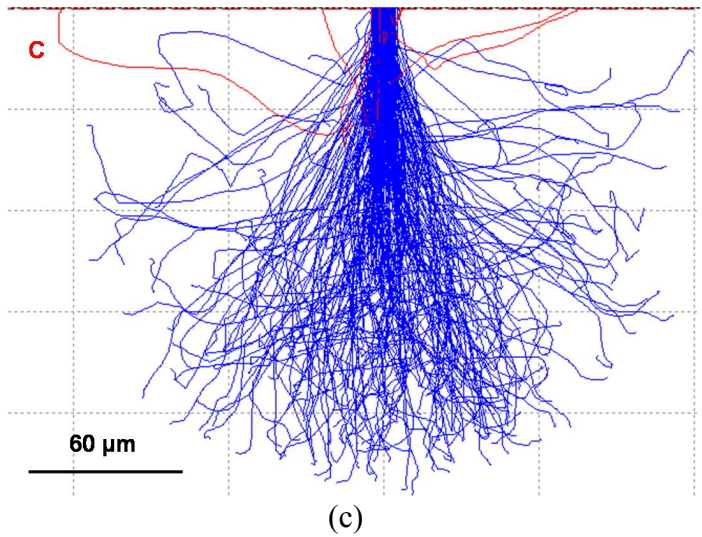
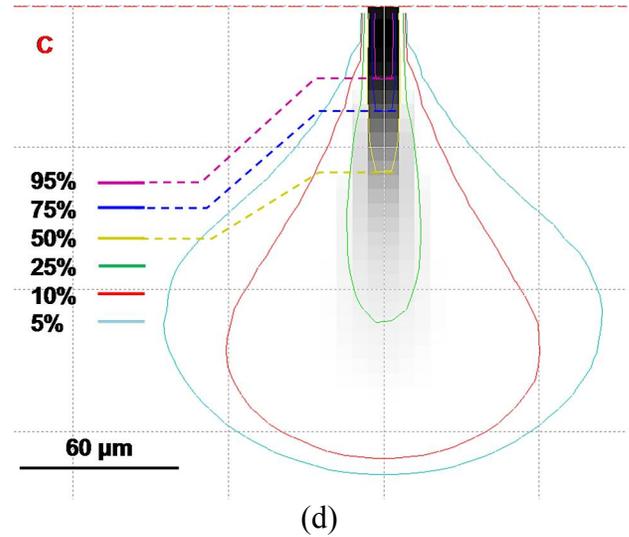

(c)　　　　　　　　　　　　　　　　　　(d)

Figure 2. (a) The NV$^-$ and NV$^0$ concentrations, as measured from fluorescence over 632–643 nm and 572–580 nm ranges, respectively, were normalized with respect to the maximum NV$^-$ concentration (N$_{max.}$) at the highest dose. (b) The locally normalized NV$^-$ and NV$^0$ concentrations are depicted as a function of depth. The depth profiles are calculated from the fluorescence spectra taken at different depths for the third highest-dose spot. (c) Electron trajectories were calculated with CASINO for 200 keV electrons incident on bulk diamond. Backscattered electrons are tracked in red. (d) The calculated electron- energy distribution.

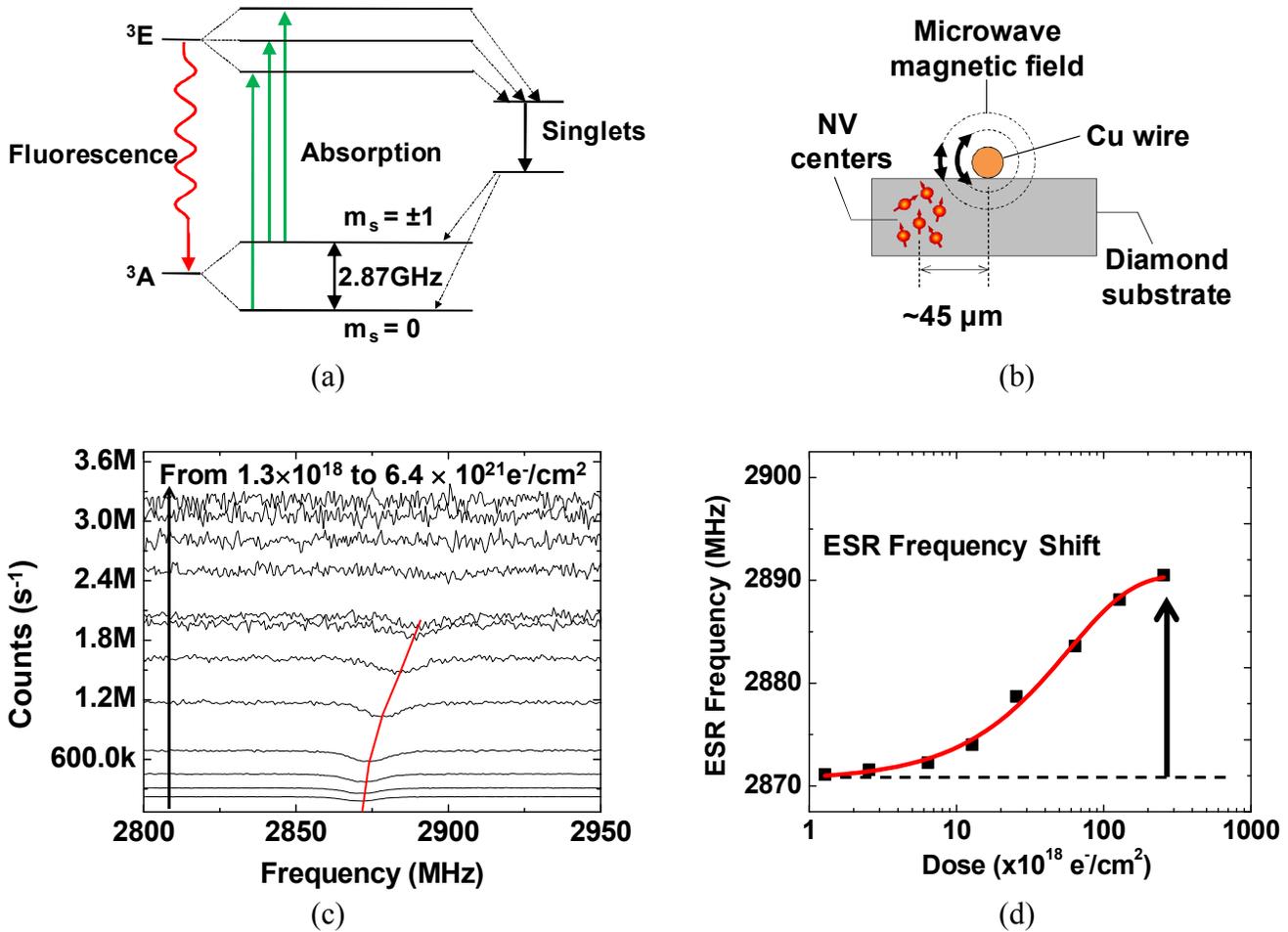

Figure 3. (a) The energy level diagram for the NV⁻ shows 2.87 GHz for the electron resonance frequency. (b) The wire is located close to the irradiated spots to apply the microwave signals and is parallel to the row of irradiated spots. (c) The CW ESR measurements are obtained using the same microwave power, $P_{mw}$ = 25 dBm. The ESR frequency for the NV center (red line) increases as the dosage becomes higher and eventually the resonance disappears. (d) The ESR frequency shifts up as the dose increases. This shift can be explained by the distortion in the diamond lattice structure caused by the electron irradiation.

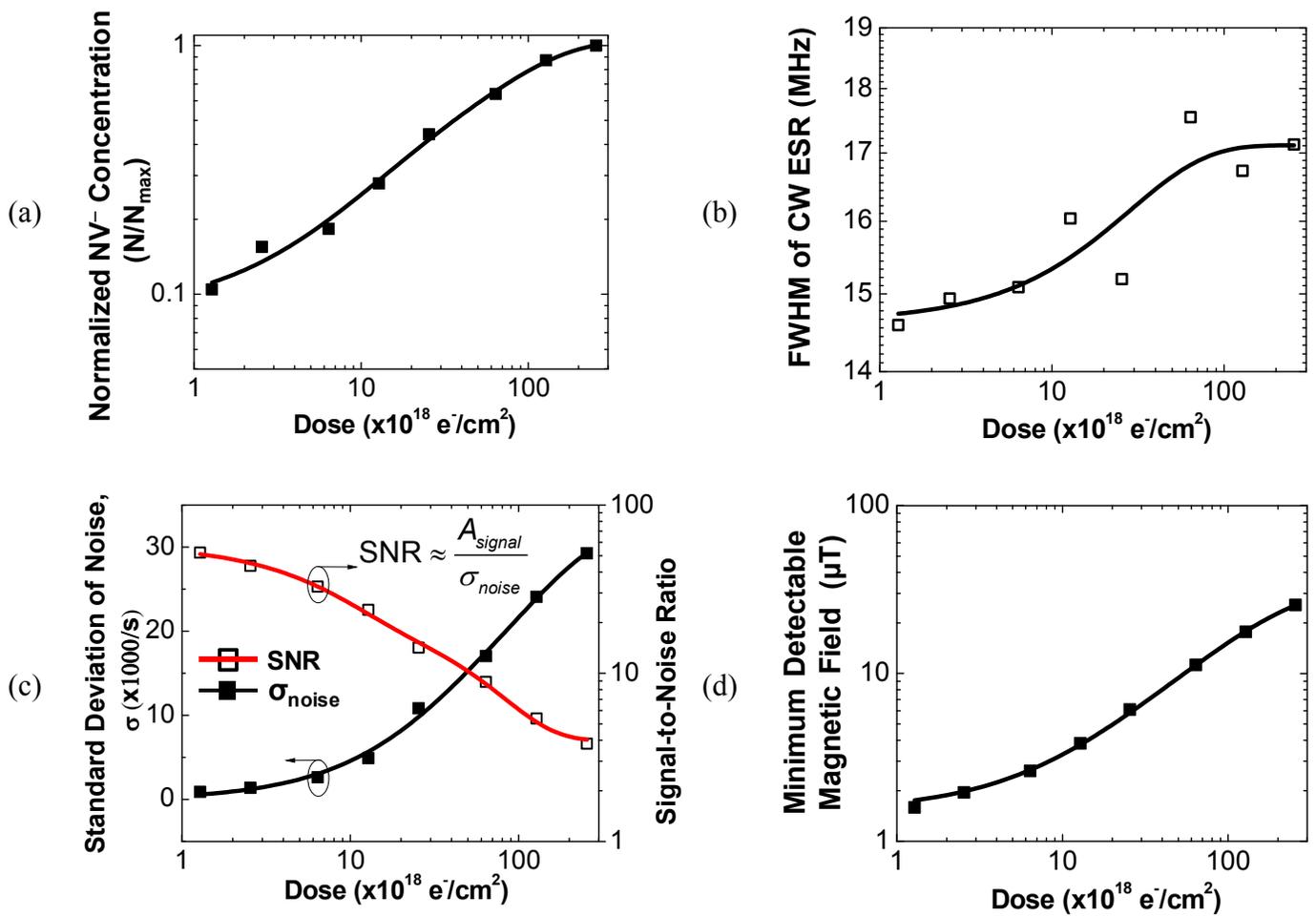

Figure 4. (a) The NV⁻ concentration is plotted vs dose for doses where the ESR signal is visible. (b) The full-width half-maximum (FWHM) of the CW ESR linewidth is also depicted. (c) The standard deviation of the noise, $\sigma_{noise}$, measured with microwave field detuned by a few linewidths from the ESR resonance frequency and the resulting signal-to-noise ratio (SNR) as a function of dose. (d) The calculated minimum detectable magnetic field with linewidth and SNR is minimized at the lowest dose.